\documentclass[a4paper,fleqn,usenatbib]{mnras}

\usepackage[T1]{fontenc}
\usepackage{ae,aecompl}
\usepackage{adjustbox}

\usepackage{graphicx}	% Including figure files
\usepackage{amsmath}	% Advanced maths commands
\usepackage{amssymb}	% Extra maths symbols

\usepackage{color}
\usepackage{float}
\usepackage{txfonts}
\usepackage{units}
\usepackage{url}
\usepackage{slantsc}

\newcommand{\lsi}{LS~I~+61$^{\circ}$303}
\newcommand{\grs}{GRS~1915+105}
\newcommand{\vcyg}{V404~Cyg}

 \newcommand{\xmm}{\textit{XMM--Newton}}
 
\title[Radio/X-ray correlations in \lsi]{Radio/X-ray correlations and variability in the X-ray binary \lsi}

\author[R.~Sharma et al.]{R.~Sharma,$^{1}$\thanks{E-mail: rsharma@mpifr-bonn.mpg.de}
 M.~Massi,$^{1}$ 
 M.~Chernyakova,$^{2,3}$
  D.~Malyshev,$^{4}$
 Y.~C.~Perrott,$^{5,6}$
 A.~Kraus$^{1}$ 
 \newauthor S.~A.~Dzib,$^{1}$
 F.~Jaron$^{7}$
and T.~M.~Cantwell$^{8}$ 
\\
$^{1}$Max-Planck-Institut f\"ur Radioastronomie, Auf dem H\"ugel 69, D-53121 Bonn, Germany \\
$^{2}$School of Physical Sciences and C-fAR, Dublin City University, Dublin 9, Ireland \\
$^{3}$Dublin Institute for Advanced Studies, 31 Fitzwilliam Place, Dublin 2, Ireland\\
$^{4}$ Institut f{\"u}r Astronomie und Astrophysik T{\"u}bingen, Universit{\"a}t T{\"u}bingen, Sand 1, D-72076 T{\"u}bingen, Germany \\
$^{5}$Astrophysics Group, Cavendish Laboratory, JJ Thomson Avenue, Cambridge CB3 0HE, UK\\
$^{6}$School of Chemical and Physical Sciences, Victoria University of Wellington, PO Box 600, Wellington 6140, New Zealand\\
$^{7}$Department of Space, Earth and Environment, Onsala Space Observatory, Chalmers University of Technology, Onsala, Sweden\\
$^{8}$JBCA, Dept. of Physics \& Astronomy, University of Manchester, Manchester M13 9PL, UK
}

\date{Accepted: 2020 November 9; in original form 2020 August 31}

\pubyear{2020}

\begin{document}
\label{firstpage}
\pagerange{\pageref{firstpage}--\pageref{lastpage}}
\maketitle

\begin{abstract}
%Context:
The high-mass X-ray binary \lsi{} exhibits variability in its radio and X-ray emissions, ranging from minute to hour time-scales. At such short time-scales, not much is known about the possible correlations between these two emissions from this source, which might offer hints to their origin.
%$Aims:$
Here, we study the relationship between these emissions using simultaneous X-ray and radio monitoring.
%$Methods and results:$ 
We present new radio observations using the Arcminute Microkelvin Imager Large Array telescope at two frequency bands, 13--15.5 and 15.5--18 GHz. We also describe new X-ray observations performed using the \xmm{} telescope. These X-ray and radio observations overlapped for five hours. 
We find for the first time that the radio and X-ray emission are correlated up to 81${{\ \rm per\ cent}}$ with their few percent variability correlated up to 40${{\ \rm per\ cent}}$. We  discuss possible physical scenarios that produces the observed correlations and  variability in the radio and X-ray emission of \lsi{}.
\end{abstract}

%%%%%%%%%%%%%%%%%%%%%%%%%%%%%%%%%%%%%%%%%%%%%%%%%%

%%%%%%%%%%%%%%%%% BODY OF PAPER %%%%%%%%%%%%%%%%%%

%%%%%%%%%%%%%%%%%%%%%%%%%%%%%%%%%%%%%%%%%%%%%%%%

\section{Introduction}

\begin{figure*}
\begin{center}
\includegraphics[width=2\columnwidth]{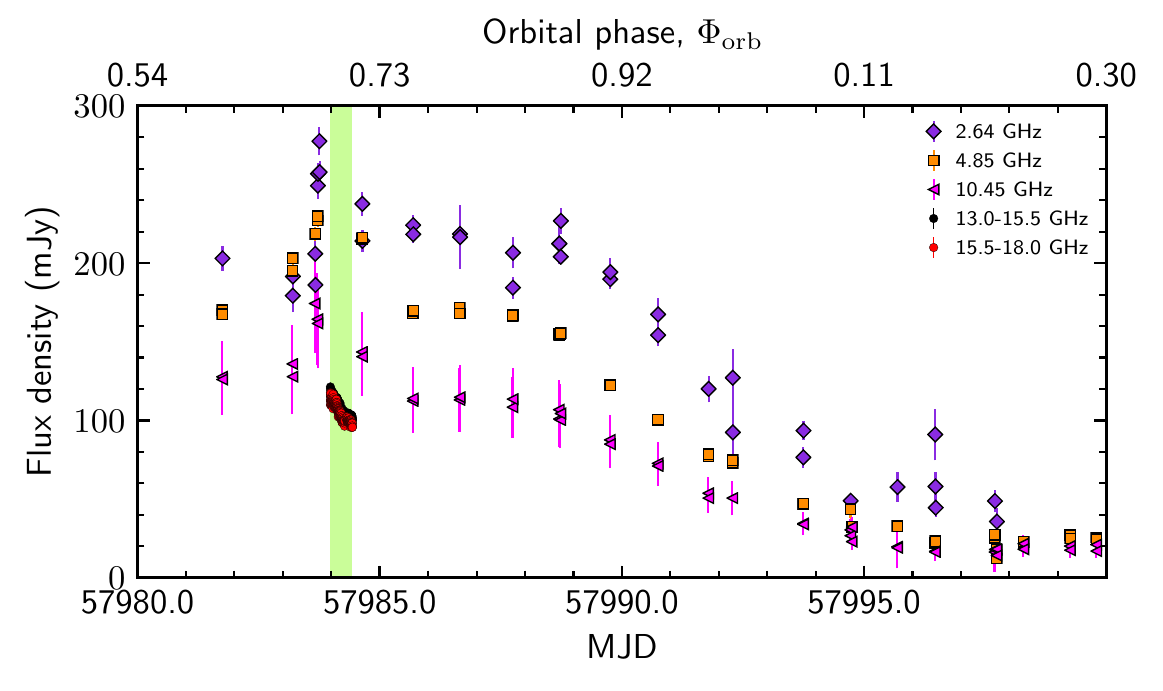}\\
\includegraphics[width=2\columnwidth]{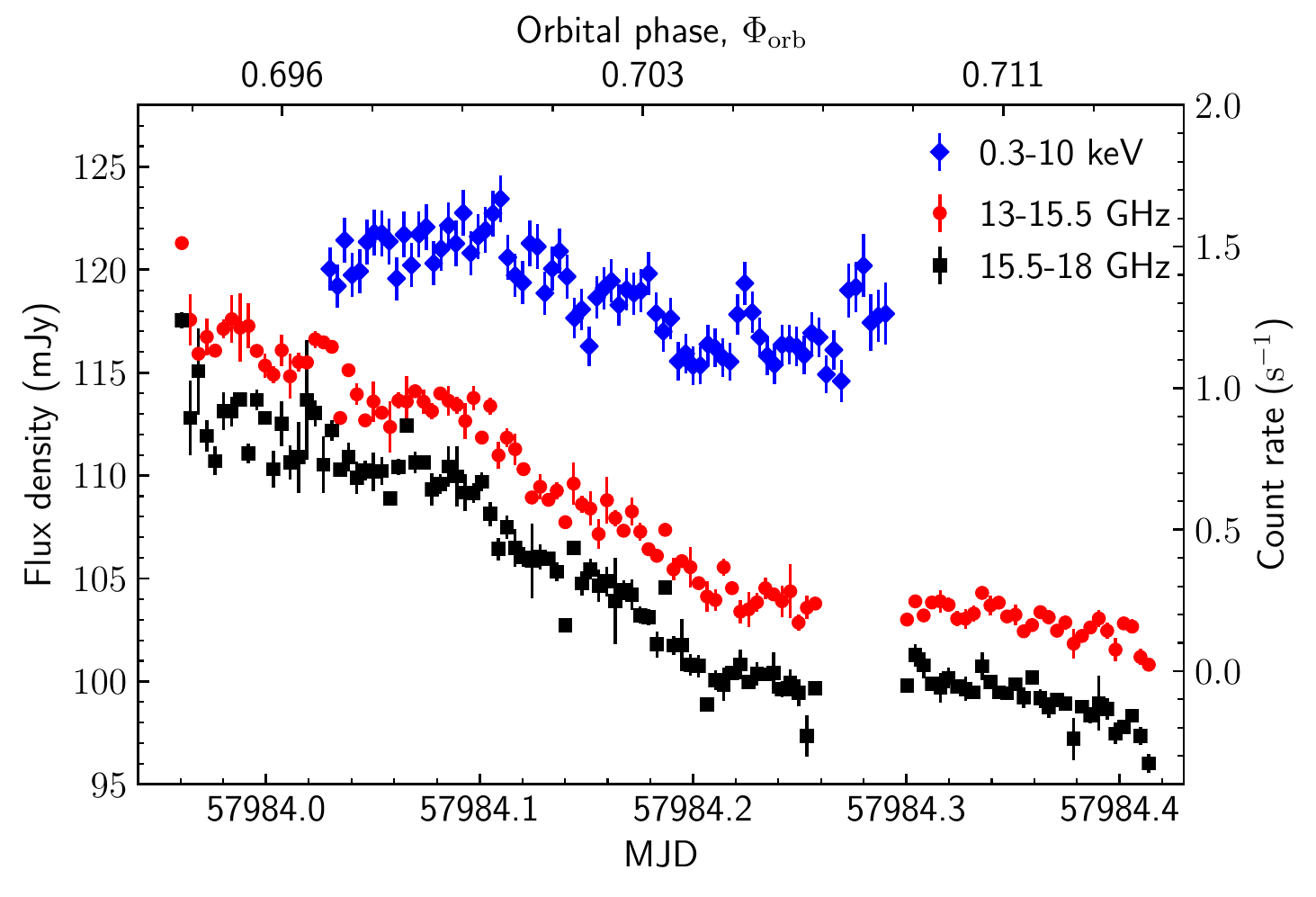}
\caption{Top: Radio light curves of \lsi{} as observed with the Effelsberg telescope at 2.64, 4.85, and 10.45 GHz (see \citealt{Massi2020}). The green-shaded region represents the time interval of observation of \lsi{} by AMI-LA telescope at 13--15.5 and 15.5--18 GHz band, which is clearly during the beginning of the decay of the radio outburst.
Bottom: Radio light curve of \lsi{} as observed with the AMI-LA telescope at frequency band of  13--15.5 and 15.5--18 GHz with units listed on the left vertical axis.
Simultaneous X-ray light curve is also shown as observed by the \xmm{} telescope in the energy range 0.3--10 keV (see right vertical axes).
Each data point corresponds to an integration time of 5~min.}
\label{fig:light curve}
\end{center}
\end{figure*}

Microquasars are X-ray binaries with emission dominated by accretion/ejection processes, observed at X-ray and radio wavelengths \citep{Mirabel1999}.
In microquasars, variability on minute--hour time-scales is investigated  
since the first observation of short-term radio variability 
in the black hole system  \vcyg{} by \citet{Han1992}.
In the black hole candidate GX~339--4, \citet{Corbel2000} reported short-term radio oscillations of $\sim$130~min at two independent radio frequencies with the higher radio frequency oscillations leading the lower frequency oscillations by few minutes. This radio emission was preceded by X-ray emission by $\sim$167~min. \citet{Tetarenko2019} showed time lags in the X-ray and radio emission of Cyg~X-1 on the order of tens of minutes. 
%More recently, \citet{Zdziarski2020}
In the well-studied source \grs{}, \citet{Fender1998} reported oscillations of period of 26~min for simultaneous radio and infrared observations with infrared emission leading the radio emission.
\citet{Mirabel1999} discussed the oscillations at different wavelengths as expanding magnetized clouds of relativistic particles. 
At high resolution, \citet{Fender1999} mapped the spatial morphology of the ejection of plasmoids in \grs{}. Between two of the mapped ejections, single dish monitoring showed quasi-periodic oscillations (QPOs) of 20--40~min. 
Plasmoids or blobs can originate in events of magnetic reconnection occurring in the jet itself (e.g.\ \citealt{Petropoulou2016,Sironi2016}) or occurring in the accretion disc \citep{Yuan2009}.
In one large reconnection region, because of instabilities such as tearing instabilities, multiple reconnection events may occur generating a sequence of flares or QPOs  \citep{Massi2011,Petropoulou2016}. 
To understand if this phenomenon is observed in the accretion disc or the jet, simultaneous X-ray and radio observations become a diagonostic tool.

\lsi{} is a radio-emitting high-mass X-ray binary with the X-ray and radio properties typical of accreting black holes with radio jet, i.e. microquasars:
the distinctive relationship between the X-ray luminosity and X-ray photon index \citep{Massi2017}, 
and the characteristic flat radio spectrum \citep{Zimmermann2015}.
In \lsi{}, \citet{Taylor1992} observed a step-like pattern in the total radio flux density  with time-scale of 1000~s. \citet{Peracaula1997} did the first timing analysis and found microflares associated with the early decay of the radio outburst with a period of 1.4~h. Longer time-scale radio QPOs of 15~h have been detected by \citet{jaron2017} at three different radio frequencies. 
In the X-ray regime, \citet{Harrison2000} observed variability on time-scales similar
to those seen in the radio by \citet{Peracaula1997}.
\citet{Chernyakova2017} reported variability time-scales of 1000~s in X-ray emission, which are similar to those observed in radio by \citet{Taylor1992}.
More recently, \citet{Nosel2018} detected QPOs for both radio ($P \sim 55$~min) and X-ray ($P \sim 2.4$~h) wavelengths, albeit at different epochs.
But the relationship between simultaneous radio and X-ray emission remains unexplored.

To probe this relationship between X-ray and radio emission in the system \lsi{}, the source was observed in 2017 August at several wavelengths for a month with a sampling rate of hours \citep{Massi2020}, 
and for a time interval of few hours, it was observed with a high sampling rate of few minutes.
In this paper we report on the radio and X-ray results of this high sampling rate monitoring, aimed to investigate the relationship and variability in the radio and X-ray emission of \lsi{}.
In Section~2, we describe the new radio and X-ray observations of \lsi{}. 
In Section~3, we describe the analysis methods employed on the data and in Section~4 we present our results. We conclude and discuss our results in Section~5.

%%%%%%%%%%%%%%%%%%%%%%%%%%%%%%%%%%%%%%%%%%%%%%%%

\section{Observations}

\lsi{} was observed extensively with the Effelsberg 100-m telescope about every 12~h from 2017 August~16 until 2017 September~13 (MJD~57981.7436 until MJD~58009.3814) at three frequencies, 2.64, 4.85, and 10.45~GHz. The light curves obtained are shown in the top panel of Fig.~\ref{fig:light curve}. The bottom axis represents time and the upper axis represents the orbital phase, $\phi _{\textrm {orb}} = \dfrac{t-t_0}{P_{\textrm {orb}}} - \textrm {int} \left(\dfrac{t-t_0}{P_{\textrm {orb}}}\right)$, where $t_0$ = MJD 43366.275 and orbital period $P_{\textrm{orb}} = 26.4960 \pm 0.0028$~d \citep{Gregory2002}. 
The eccentricity of the source has not been firmly established (see \citealt{Kravtsov2020}).
Extensive study on this data is presented in \citet{Massi2020}. Below is the description of the radio and X-ray observations performed at a higher sampling rate.

\subsection{AMI-LA}

Radio observations of the source were performed with the AMI-LA \citep{2008MNRAS.391.1545Z} using the upgraded correlator as described in \citet{2018MNRAS.475.5677H}. 
The observations were performed continuously for almost 11~h from 2017 August 18--19  (MJD~57983.9823 until MJD~57984.4362). The observations were performed about every 1.9 min. The bright point-like source J0228+6721 was observed in an interleaved fashion every 700\,s throughout the observation. The observation was carried out at two frequency bands: 13--15.5 and 15.5--18~GHz.

Data reduction was performed using \textsc{\texttt{casa}}\footnote{\url{https://casa.nrao.edu/}} \citep{McMullin2007}.  The standard calibration source 3C\,286 was used to set the flux density scale, using the \citet{2013ApJS..204...19P} flux scale together with a correction for the $I+Q$ polarization measured by AMI-LA derived from the polarization measurements from \citet{2013ApJS..206...16P}. The interleaved calibration source J0228+6721 was used for phase calibration, and \lsi{} was also self-calibrated in phase assuming a point source model, adequate for the $\approx$\,30\,arcsec resolution AMI-LA data. An amplitude correction for atmospheric emission and absorption calculated using the `rain gauge' noise injection system was also applied.  After these calibrations, a small spectral index deviation correlating with a change in atmospheric properties was observed on both the calibrator and the source, and was therefore attributed to a change in atmospheric/instrumental properties and this interval between MJD~57984.2772  and MJD~57984.3200 was removed. For comparing with the \xmm{} observations, as described in the next section, the 5~min integrated light curves are shown in the bottom panel of Fig.~\ref{fig:light curve} (the red circles and the black squares). The errors reflect the standard deviation. Correlations and timing analysis described in Sections~3 and~4 were performed either on the two independent radio bands or on the flux obtained by fitting over the entire band in order to improve the signal-to-noise ratio.

\subsection{\xmm}

In this work, we use \xmm{} observation (Obs. Id. 0795711501) of almost 6~h taken on 2017 August~19. The X-ray data are simultaneous with the radio data for $\sim$~5~h. To analyse it, we used the \xmm{} \texttt{Science Analysis software\footnote{\url{https://www.cosmos.esa.int/web/xmm-newton/what-is-sas}} v.18.0.0}. Known hot
pixels and electronic noise were removed, and data were filtered to exclude episodes of soft proton flares. The total clean exposure is $\sim 23$~ks.  The light curve was extracted for EPIC-pn camera from a $40$ arcsec radius circle centred at the position of \lsi{} and the background was extracted from a nearby source-free region of $80$ arcsec radius. 
The obtained light curve in the 0.3--10~keV energy range with 5~min time binning is shown in the bottom panel of Fig.~\ref{fig:light curve} (the blue diamonds).

\begin{figure}
\begin{center}
\includegraphics[width=1\columnwidth]{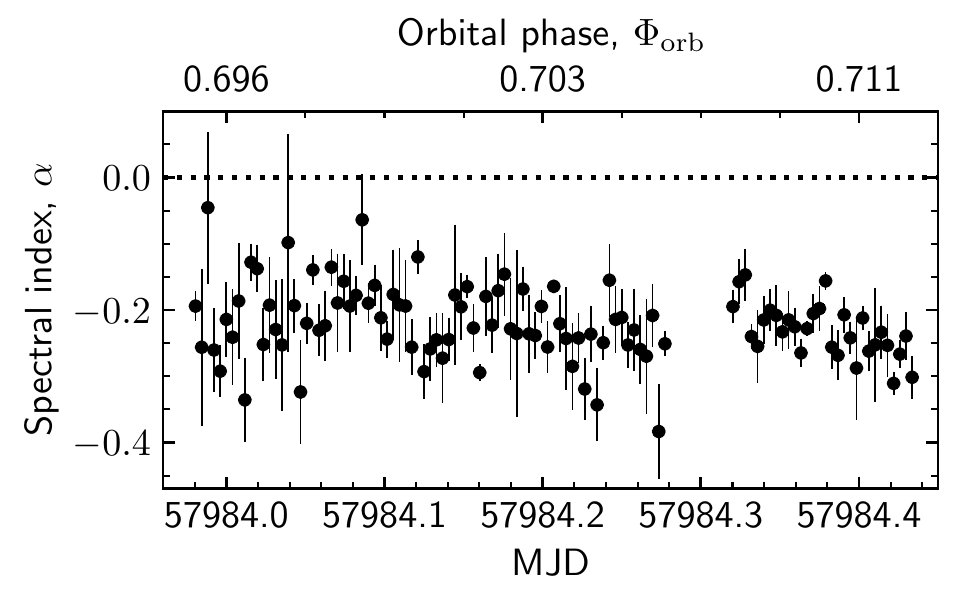} 
\caption{Spectral index variations with time for radio data of \lsi{} obtained with the AMI-LA telescope and integrated every 5~min.}
\label{fig:spectral_index}
\end{center}
\end{figure}

\begin{figure}
\begin{center}
\includegraphics[width=1\columnwidth]{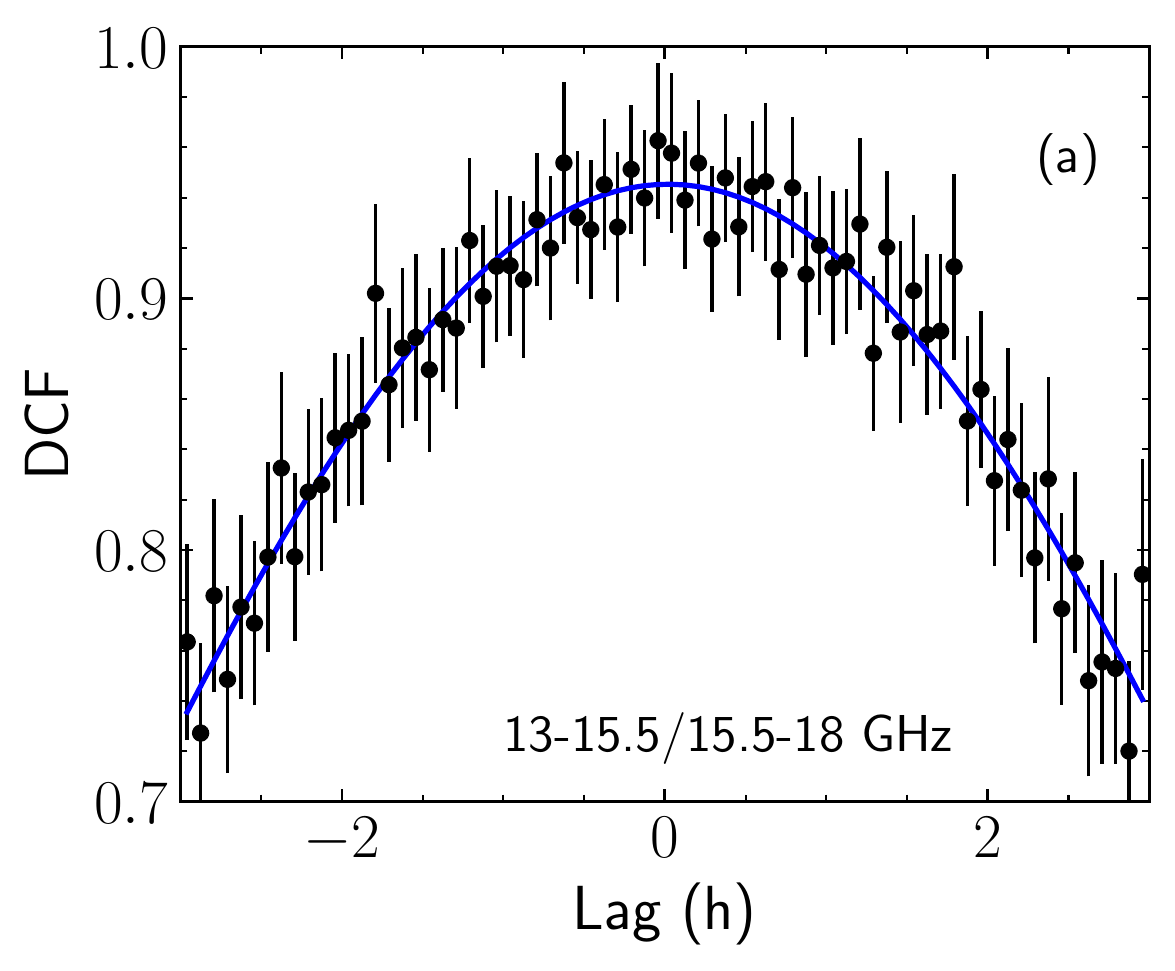} \\
\includegraphics[width=1\columnwidth]{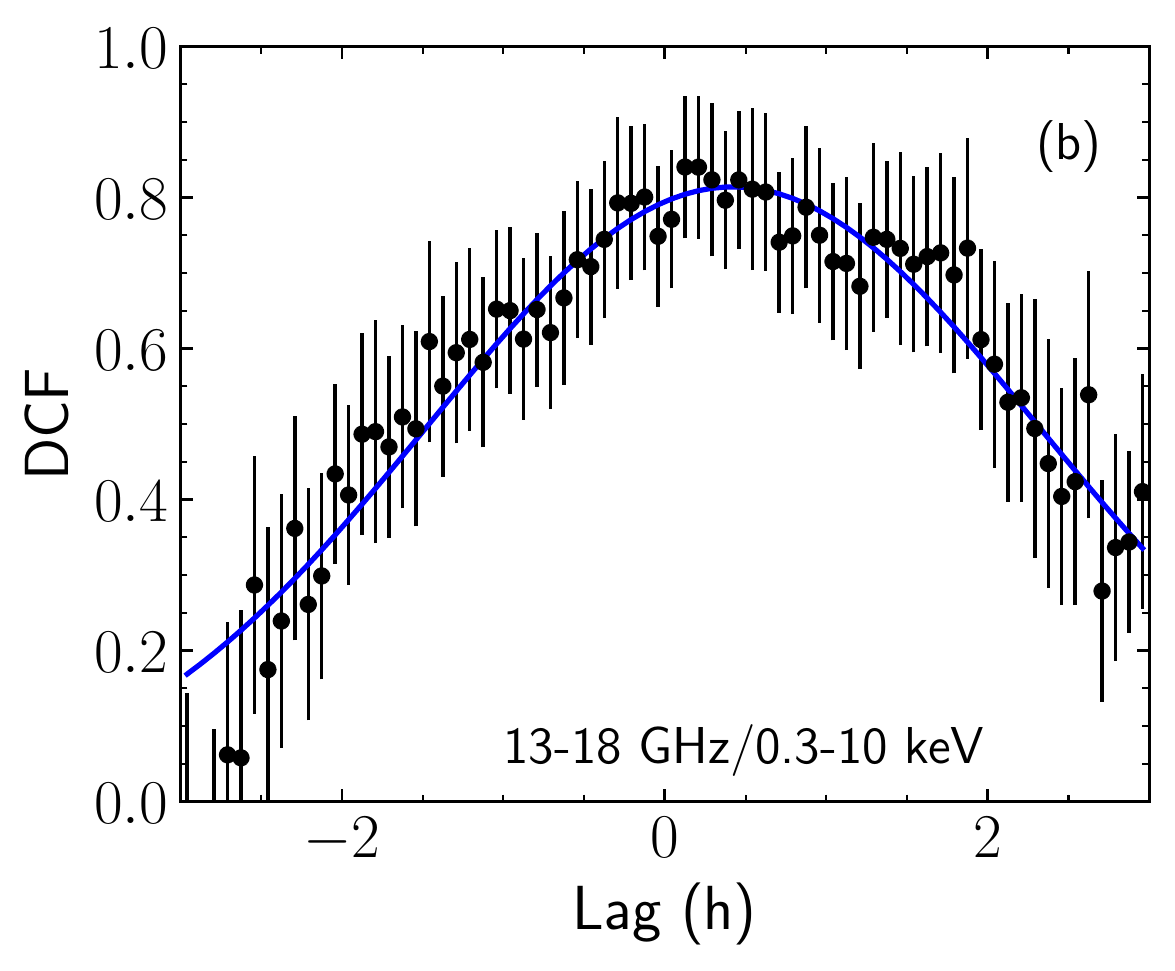}
\caption{Discrete cross-correlation function of the total flux of (a) 15.5--18 GHz and 13--15.5 GHz and  (b) 13--18 GHz/0.3--10 keV. The blue curves represent the Gaussian fit to the data with fit parameters given in Table~\ref{table:fit_gaussian}. Positive time lag hints that the radio emission precedes the X-ray emission but the lag is consistent with zero  within Gaussian dispersion.}
\label{fig:DCF_tot}
\end{center}
\end{figure}

\section{Data analysis} 

To study the relationship between the light curves, we analysed them using the discrete correlation function  (DCF;   \citealt{Edelson1988, Hufnagel1992}). This method is used to find correlations and possible time lags in unevenly sampled light curves that is usually the case for astronomical observations and it provides meaningful uncertainty to the correlation coefficients (e.g. \citealt{Fender2002, Chidiac2016, Alvarez2018}). Given a time series, $a$ and $b$ with $a(i)$ and $b(j)$ being the individual data points of the time-series, initially the unbinned discrete correlation function (UDCF) is calculated for all possible time lags, $\Delta t_{ij} = t_{j}-t_{i}$ as
\begin{equation}
\mbox{UDCF}_{ij} = \dfrac{(a(i)-\overline{a})(b(j)-\overline{b})}{\sqrt{\sigma_{a}^{2} \sigma_{b}^{2} }},
\end{equation}
where $\bar{a}$ and $\bar{b}$ are the respective mean of the time series and $\sigma_{a}^{2}$ and $\sigma_{b}^{2}$ are their variance.
The UDCF is then binned in time taking $\tau$ as the centre of the bin and $M$ as the number of pairs in each bin,
\begin{equation}
\mbox{DCF} (\tau) = \dfrac{1}{M} \sum_{\tau - \Delta \tau/2}^{\tau + \Delta \tau/2} \textrm{UDCF}_{ij} (\tau).
\end{equation}
The error for each bin is given by
\begin{equation}
\sigma_{DCF} (\tau) = \dfrac{1}{M-1} \sqrt{\sum_{\tau - \Delta \tau/2}^{\tau + \Delta \tau/2} \Big[\textrm{UDCF}_{ij} - \textrm{DCF} (\tau)\Big]^2 }.
\end{equation}
The normalisation (mean and standard deviation of respective light curves) assumes that the light curves are statistically stationary, which is not true for variable data. To make the light curve stationary, the mean and standard deviation are calculated using only the points that overlap in a given time-lag bin, thus making the DCF  bound between [$-1, +1$] interval \citep{White1994}.
Also as mentioned in \citet{Edelson1988}, to keep the normalisation correct, we need to omit the zero lag pairs (i.e. pairs where $i=j$).
For a more quantitative study, we performed least-squares fit of a Gaussian to the DCF profiles.
Thus, to estimate the time-lag between emission at two frequencies, a Gaussian function of the following form is fit to the DCF:
\begin{equation}
\mbox{DCF}(t) = A \times \mbox{exp}\Bigg[\dfrac{-(t-T)^2}{2w^2}\Bigg],
\label{eq:gaussian}
\end{equation}
where $A$ is the peak DCF value, $T$ is the time-lag at peak of DCF, and $w$ is the width of the Gaussian function.
As a conservative approach we use the width, $w$  as an error on the time-lag (see \citealt{Alexander2013, Chidiac2016}).

In Fig.~\ref{fig:light curve} (bottom panel), we note that there are small variability of few mJy and counts s$^{-1}$ superimposed over the long-term decaying trend of the radio and X-ray light curves, respectively. The analysis of these small variations require the removal of the long-term trend  from the light curves (e.g. see \citealt{Fender2002, jaron2017}).
In order to remove the long-term trend from the data, we subtraced a piecewise linear function from the data.
For the radio light curves, we fit the linear function first from MJD 57983.9823 to MJD 57984.2772 and then from MJD 57984.3200 to MJD 57984.4362.
For the XMM data, the fit was applied to the whole interval.

\begin{figure*}
\begin{center}
\includegraphics[width=2\columnwidth]{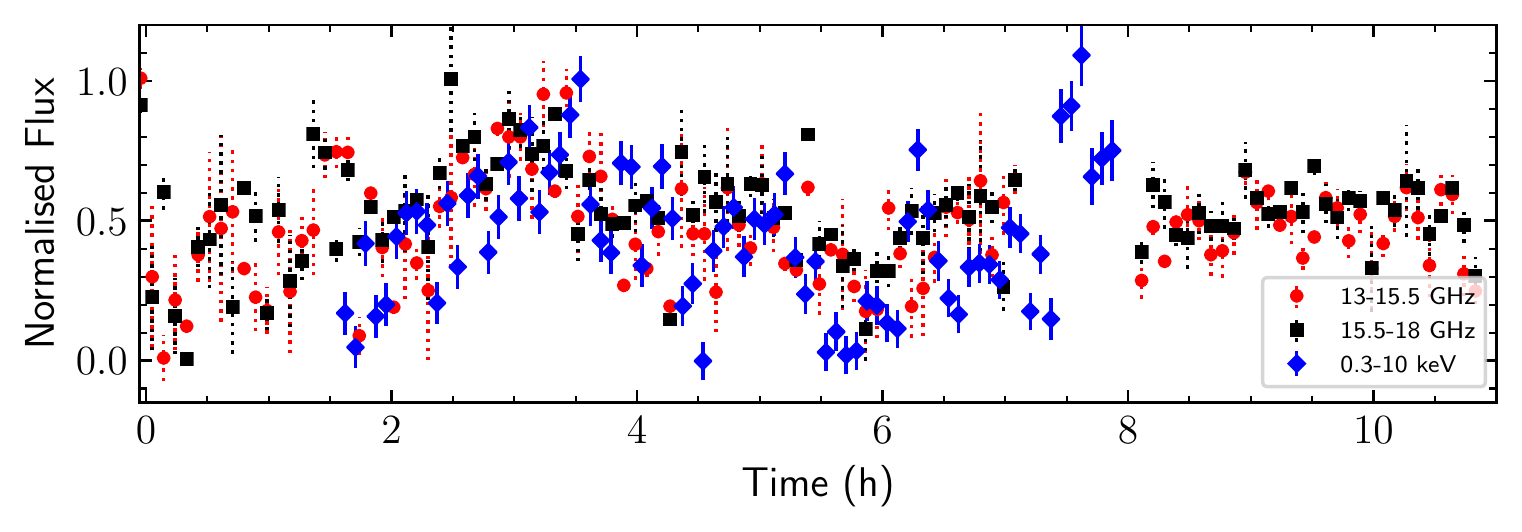} \\
\includegraphics[width=2\columnwidth]{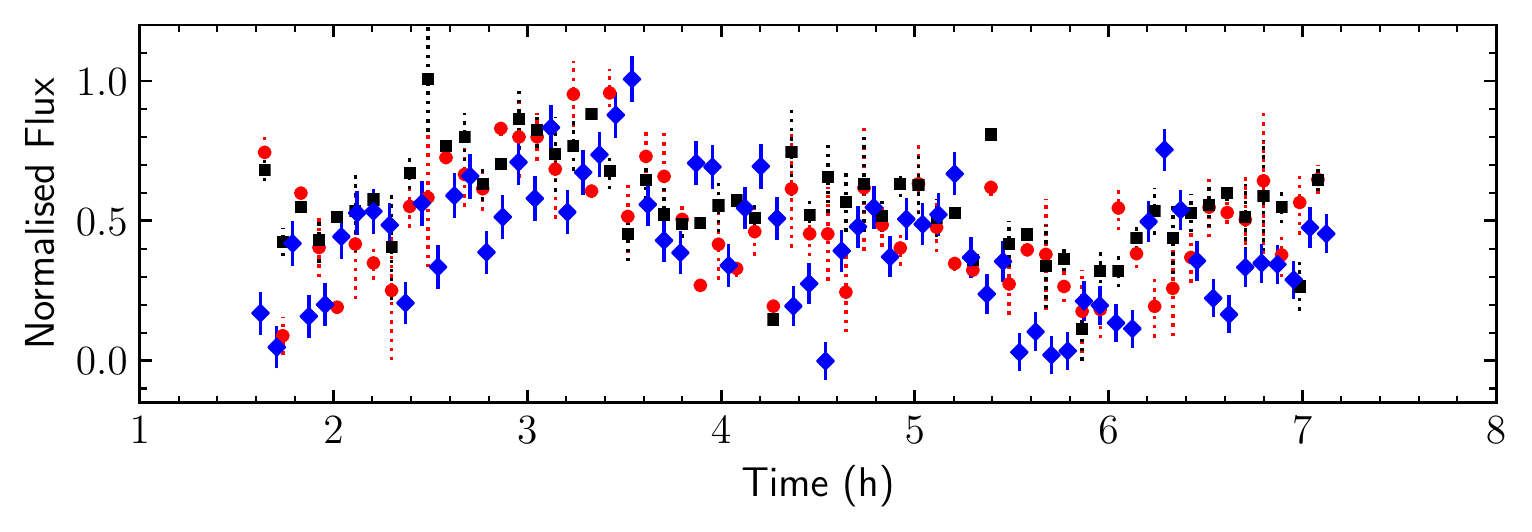} \\
\caption{Detrended and normalized light curves for comparison. Top: For the entire duration of the observations of about 11~h with 13--15.5 GHz, 15.5--18 GHz, and 0.3--10 keV represented by the red circles, the black squares, and the blue diamonds, respectively.  Bottom: Zoom-in of the light curves only for the simultaneous radio and X-ray observations of 5~h.}
\label{fig:overlap}
\end{center}
\end{figure*}

\setcounter{table}{0}
\begin{table}
\caption{Best-fitting parameters of the Gaussian function fitted to the DCF in Fig.~\ref{fig:DCF_tot} according to Eq.~\ref{eq:gaussian}. $A$ is the amplitude of the Gaussian, $T$ is the time-lag, and $w$ represents the width. The error in brackets of time-lag $T$ represent the fit errors. Although using a conservative approach, the error on time-lag is given by the Gaussian width, $w$.}
\centering
\begin{tabular}{lcccc}
\hline
Frequency  & $A$  & $T$ (min) & $w$ (h) & $\chi_{\textrm{red}}^2$   \\ \hline 
13--15.5/15.5--18 GHz &  0.94 $\pm$ 0.01 &  1.2 [$\pm$ 1.2]   & 4.0 $\pm$ 0.1 &  0.2  \\ 
13--18 GHz/0.3--10 keV & 0.81 $\pm$ 0.01 & 25.2 [$\pm$ 2.4]   & 1.9 $\pm$ 0.1  & 0.3   \\ 
\hline
\end{tabular}
\label{table:fit_gaussian}
\end{table}

\begin{figure}
\begin{center}
\includegraphics[width=1\columnwidth]{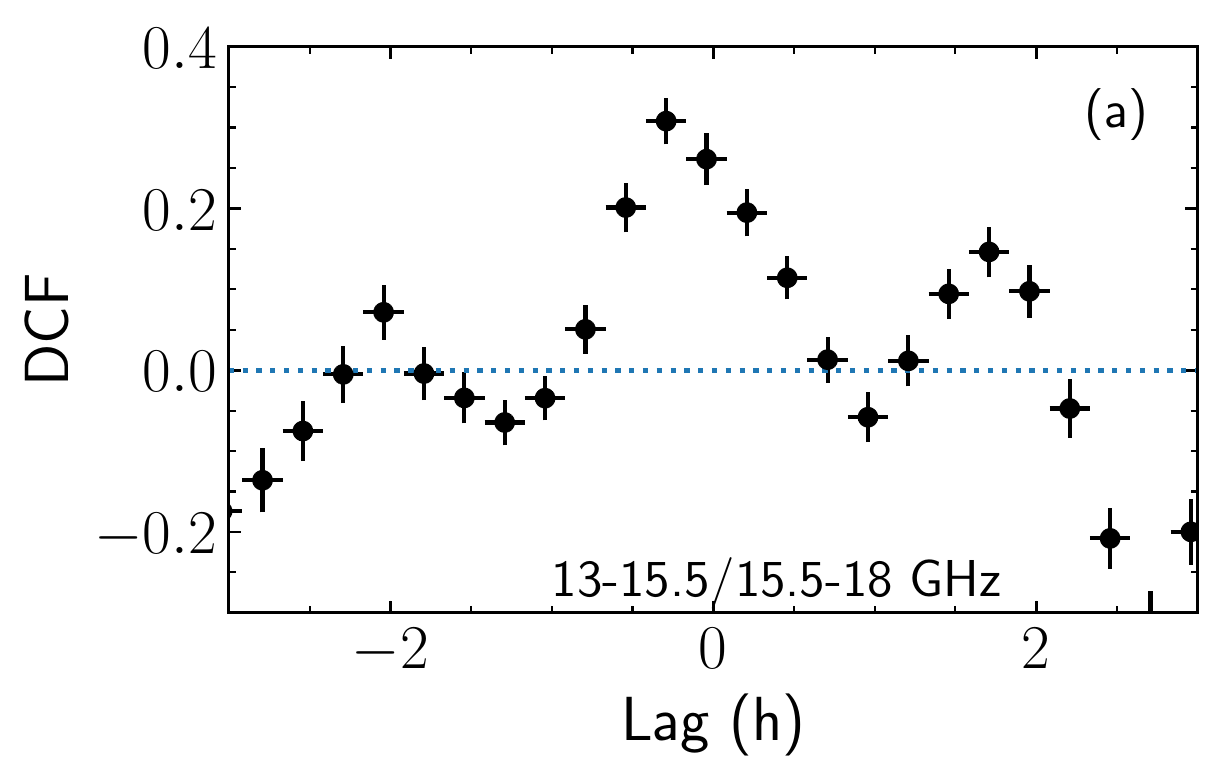}\\
\includegraphics[width=1\columnwidth]{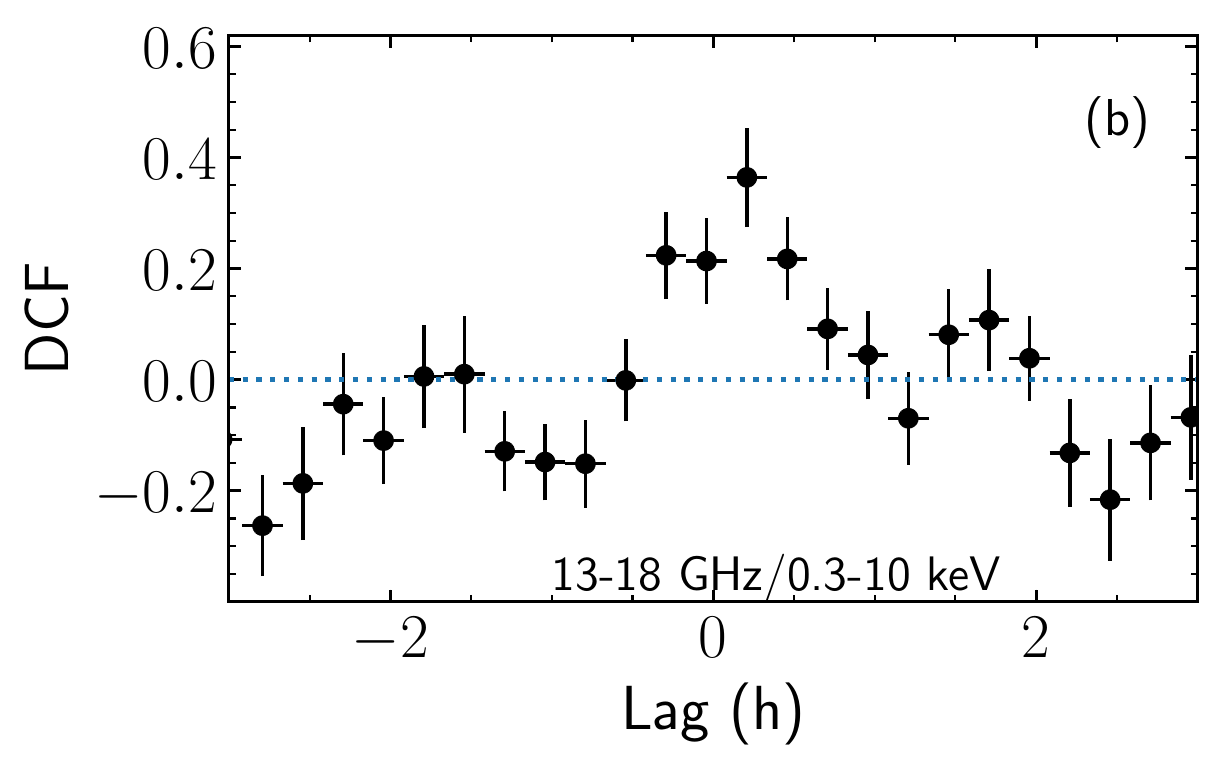}\\
\includegraphics[width=1\columnwidth]{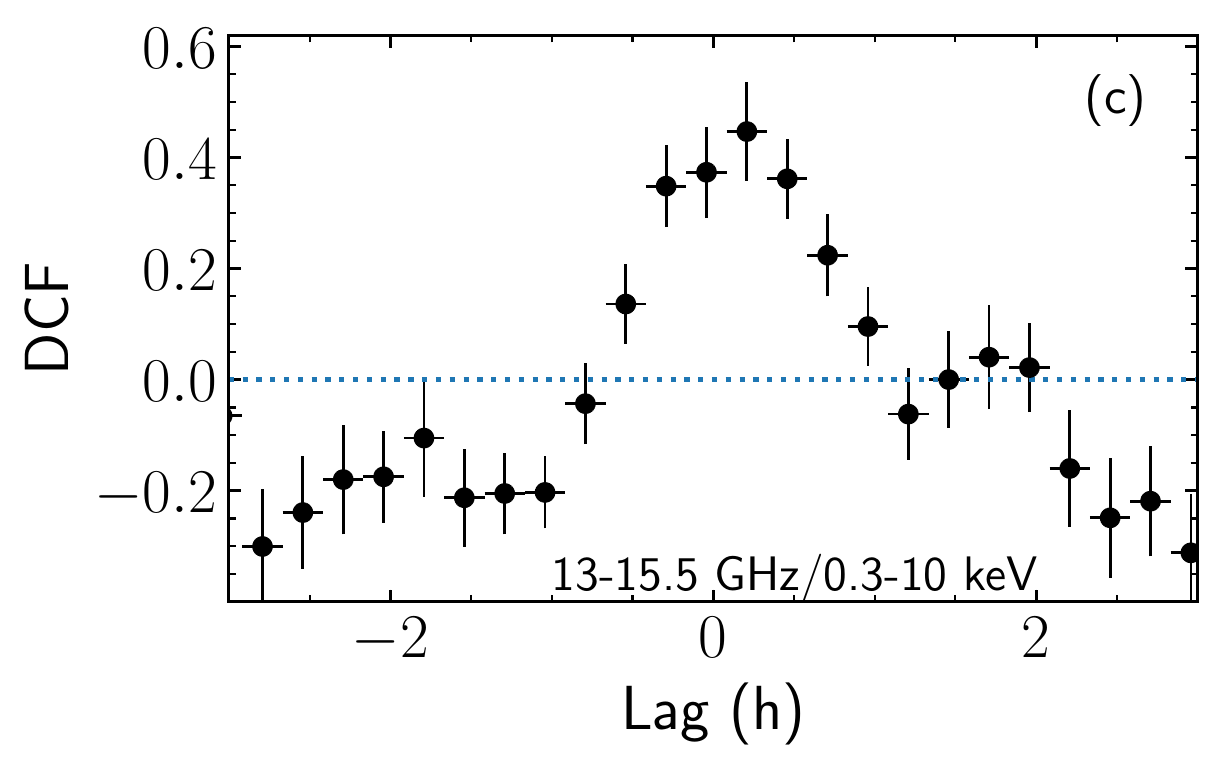}\\
\includegraphics[width=1\columnwidth]{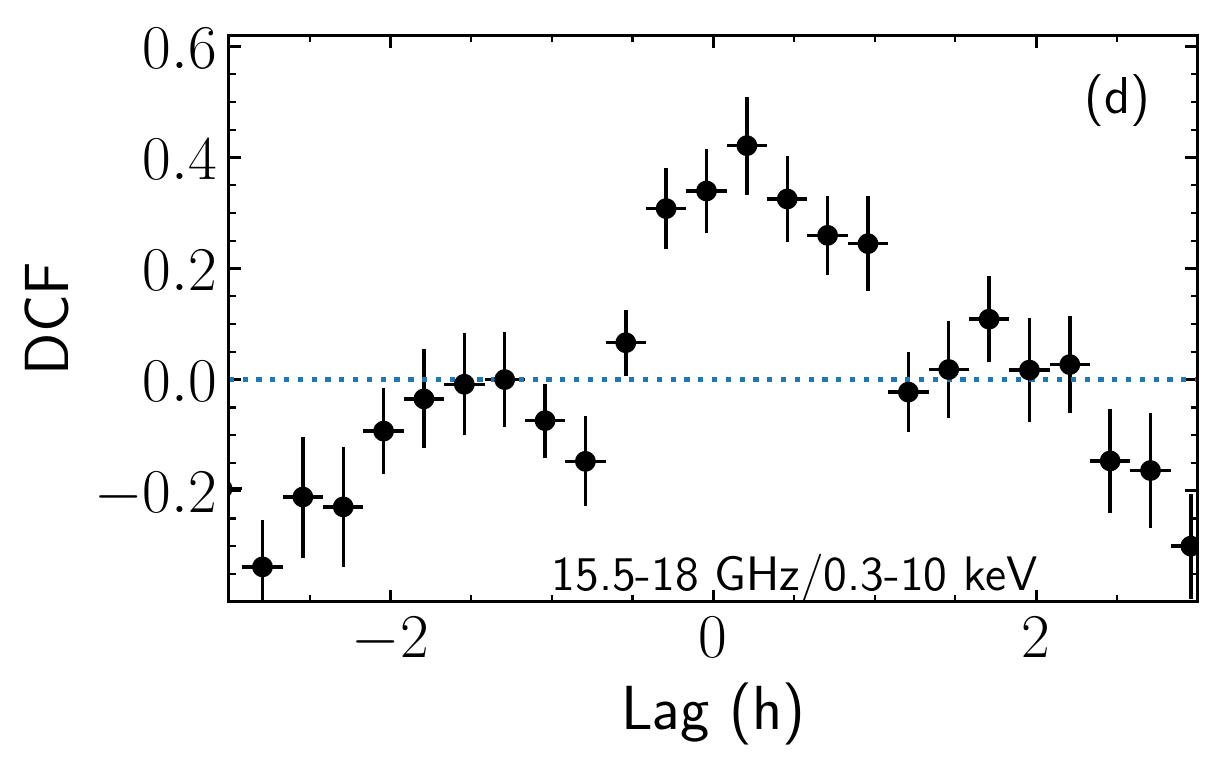}
\caption{Discrete cross-correlation of detrended light curves of (a) 15.5--18 and 13--15.5 GHz with secondary peaks representing possible periodicity, (b) 13--18 GHz/0.3--10 keV, (c) 13--15.5 GHz/0.3--10 keV, and (d) 15.5--18 GHz/0.3--10 keV. All correlations peak around zero time-lag. The blue-dotted line represents correlation coefficient of zero.}
\label{fig:DCF_rect}
\end{center}
\end{figure}

\begin{figure}
\begin{center}
\includegraphics[width=1\columnwidth]{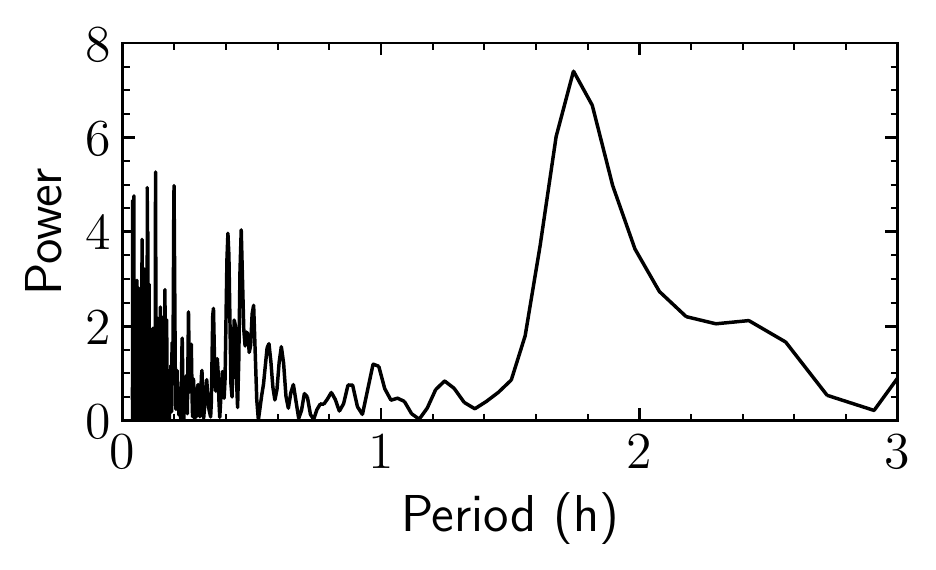}\\
\includegraphics[width=1\columnwidth]{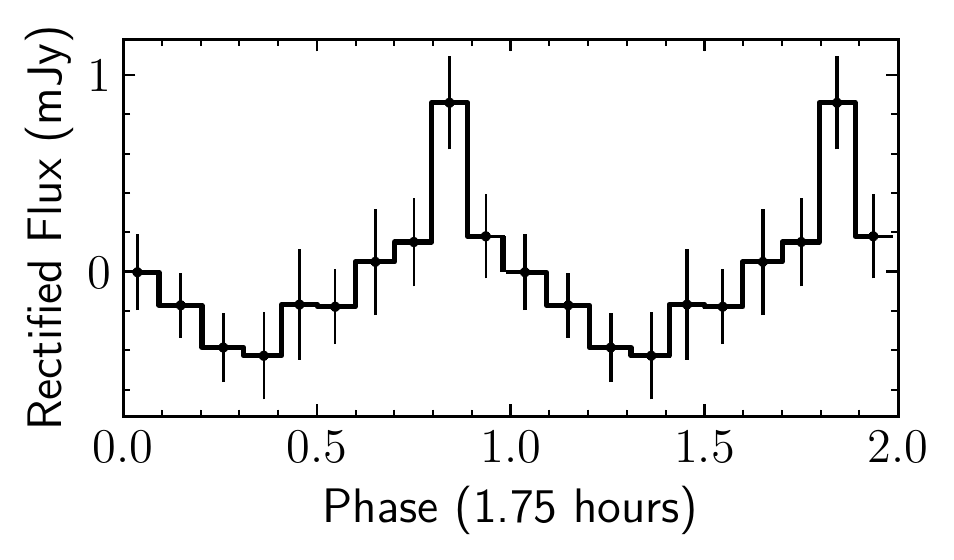}\\
\includegraphics[width=1\columnwidth]{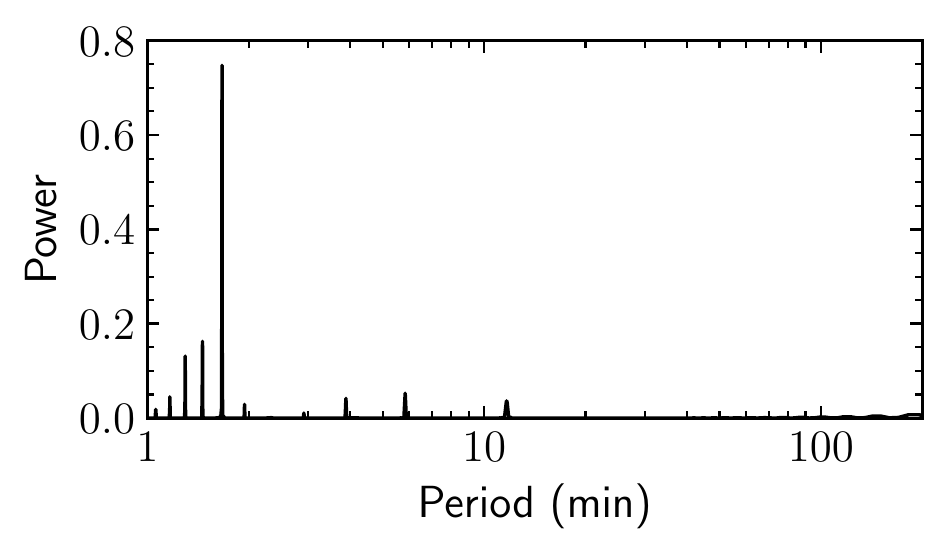} 
\caption{Top: Lomb--Scargle periodogram of 13--18 GHz detrended radio data showing a significant peak at 1.75 $\pm$ 0.10~h. Middle: Phase-averaged data folded with the significant period of 1.75 h in 10 bins. For clarity, the data are repeated in the second cycle. Bottom: Fourier transform of the sampling function of radio data showing periodic feature with maximum power at a period of 1.7~min and no feature at 105~min (1.75~h). The x-axis is in log scale.}
\label{fig:timing_analysis_AMI-LA_13GHz}
\end{center}
\end{figure}

In order to find the statistical significance of the variability, we calculate the fractional variability \citep{Vaughan2003} of the detrended data as
\begin{equation}
F_{\textrm{var}} = \sqrt{ \dfrac{S^2- \overline{\sigma^2_{\textrm{err}}}}{\overline{x}^2}},  
\end{equation}
where $S^2$ is the variance of the light curve, $\overline{\sigma^2_{\textrm{err}}}$ is the mean square error, and $\overline{x}^2$ is the mean flux of the light curve. The error in the fractional variability is estimated as
\begin{equation}
\textrm{err}(F_{\textrm{var}}) = \sqrt{ \dfrac{1}{2N} \left(\dfrac{\overline{\sigma^2_{\textrm{err}}}}{F_{\textrm{var}} \times \overline{x}^2 }\right)^2   +  \dfrac{\overline{\sigma^2_{\textrm{err}}}}{N \times \overline{x}^2}}. 
\end{equation}
The detrended data were then analysed using the Lomb--Scargle periodogram, which is a powerful tool to find and test the significance of weak periodic signals in unevenly sampled data \citep{Lomb1976, Scargle1982}.
Fischer-randomization test is used to test the significance of the found periodic signal,
where the flux is permuted thousand times and thousand new
randomized time series are created and their periodograms calculated \citep{Nemec1985}.
The proportion of permuted time series that contain a higher peak in the periodogram than the original
periodogram at any frequency then gives the false alarm probability, $p$ of the peak. If $p < 0.01$,
the period is significant and if $0.01 < p < 0.1$ the period is marginally significant.
In order to check if the found period is a true period and it is not generated due to sampling, the sampling function is analysed. The flux density are assigned a constant value equal to one. This function is further analysed using the Fourier transform.

\section{Results}

The radio and X-ray light curves are shown in the bottom panel of Fig.~\ref{fig:light curve}. We see that the radio emission displays a decaying trend with the flux density decreasing from around 120~mJy to 95~mJy in $\sim$11~h observation. In fact, as shown in the top panel of Fig.~\ref{fig:light curve}, 
the observation with the AMI-LA telescope took place 
during the beginning of the decay of the large radio outburst of \lsi{}. The X-ray light curve also shows a general decaying trend in its counts s$^{-1}$. Figure~\ref{fig:spectral_index} shows the radio spectral index, $\alpha = \log(S_1/S_2)/\log(\nu_1/\nu_2)$, which is negative throughout the 11~h observation.
In this section, we first discuss the results of the correlation analysis of the simultaneous radio and X-ray data and then of the variability. 

To find the strength of association between the two time series, we calculated the Spearman correlation ($r_s$).
For 13--15.5/15.5--18~GHz, $r_s = 0.94$, for  13--15.5 GHz/0.3--10~keV, $r_s = 0.84$ and for 15.5--18 GHz/0.3--10 keV, $r_s = 0.83$. All correlations have a p-value$< $0.001 which means that it is very unlikely to get this result by chance. Squaring the coefficients for the radio/X-ray data yields the strength of correlation $r_\mathrm{ s}^2\sim 0.7$, i.e. 70{{\ \rm per\ cent}} of the variability in one frequency is due to the variability in the other frequency.

To further quantify this, we also compute the discrete cross-correlation functions (DCF).
Figure~\ref{fig:DCF_tot}(a) shows the cross-correlation between the two radio bands, i.e. 13--15.5 and 15.5--18 GHz. The blue curve represents the Gaussian fit to the data with fit parameters given in Table~\ref{table:fit_gaussian}. We see that the two frequency bands are strongly correlated up to 94${{\ \rm per\ cent}}$ with zero time-lag.
Cross-correlations between simultaneous radio and X-ray data are shown in Fig.~\ref{fig:DCF_tot}(b). We find that the radio band is strongly correlated with the X-ray. The DCF peaks at $81 {{\ \rm per\ cent}}$ with a time-lag of $\sim$25~min and  a conservative error estimate of $\sim$ 2 h.

To further analyse the data, we work with the detrended light curves. The fractional variability amplitude of the radio frequencies is $\sim 64.0 \pm 0.2 {{\ \rm per\ cent}}$ and that of the X-ray is $\sim 15.0 \pm 5.0 {{\ \rm per\ cent}}$.
The normalized detrended flux for both the radio and X-ray emission (normalized with respect to the maximum flux density in their respective light curves) are shown in the top panel of Fig.~\ref{fig:overlap} and a zoom-in of the same for the 5~h simultaneous observations is shown in the bottom panel of Fig.~\ref{fig:overlap}.
There is  variability  with  varying amplitude. 
Qualitatively, the variability at both radio frequencies seems to coincide with the variability in the X-rays (see Fig.~\ref{fig:overlap}). 
We perform the cross-correlation analysis using the simultaneous X-ray and radio data of the detrended light curves. 
Figure~\ref{fig:DCF_rect}(a) shows the cross-correlation between the two radio bands. The DCF peaks at around zero time lag with a coefficient of 0.30 $\pm$ 0.04.
There is also a correlation at negative time lag of 2.0 $\pm$ 0.1~h and positive time lag of 1.7 $\pm$ 0.1~h with DCF coefficients of 0.10 $\pm$ 0.04  and 0.13 $\pm$ 0.04, respectively.
These secondary peaks hint 
at a possible periodicity in the radio data.
The cross-correlation between the detrended radio and X-ray data is shown in Fig.~\ref{fig:DCF_rect}(b).
The DCF of radio band with X-ray peaks around zero lag with a coefficient of 0.36 $\pm$ 0.10. The same DCF pattern is also evident when we correlate the X-ray data with the two independent radio bands (Figs~\ref{fig:DCF_rect}c and d). The DCF of both radio bands with X-ray peaks around zero lag with a coefficient of 0.45 $\pm$ 0.10 and 0.42 $\pm$ 0.10 for 13--15.5 GHz/0.3--10 keV and 15.5--18 GHz/0.3--10 keV, respectively.

As mentioned earlier, the radio and X-ray light curves show variability.
The 5~h simultaneous observations of radio and X-ray data do not reveal significant features in the Lomb--Scargle periodogram. Therefore, we further use the full 11~h radio data. Top panel of Fig.~\ref{fig:timing_analysis_AMI-LA_13GHz} shows the Lomb--Scargle periodogram for 13--18~GHz radio band, with a dominant and significant peak at 1.75 $\pm$ 0.10~h, corresponding to the secondary peaks in the DCFs. The  phase-averaged data folded with a period of 1.75 h are shown in the middle panel of Fig.~\ref{fig:timing_analysis_AMI-LA_13GHz}.
As shown in the bottom panel of Fig.~\ref{fig:timing_analysis_AMI-LA_13GHz}, we see a dominant feature in the sampling function at about 1.7~min, i.e. the sampling rate.
No feature is present at the found period of 105~min (i.e. 1.75~h) due to which we know that the found period is not an effect of the sampling. This confirms that the radio frequency has a QPO of 1.75 $\pm$ 0.10~h.

\section{Conclusions and discussion}
We observed \lsi{} with the AMI-LA telescope  for around 11~h at two radio frequency bands, 13--15.5 and 15.5--18 GHz. 
We also observed it with the \xmm{} telescope for about 6~h at energy range of 0.3--10 keV, 5~h of which was simultaneous with the radio observations. Both radio and X-ray show variability superimposed over the long-term trend.

We find that the radio and X-ray emission are correlated up to 81${{\ \rm per\ cent}}$. 
In the detrended light curves, i.e. after the removal of the long-term trend, the variability in X-ray and radio emission  are correlated up to 40${{\ \rm per\ cent}}$.  
The radio variability overlaps with the X-ray variability. The DCF analysis hints that the radio emission leads the X-ray emission by $\sim$25~min, but within Gaussian dispersion they have zero time-lag.
The better signal-to-noise ratio and longer observation time interval of radio data allow us to  establish by
Lomb--Scargle periodogram QPOs with a period of 1.75~$\pm$~0.10~h in \lsi{}.

The significant radio/X-ray correlation implies that both the radio and X-ray emission are due to the same electron population.
But are these emissions due to the same or different physical processes?
Our simultaneous radio/X-ray observations were performed  during the beginning of the decay of a major  radio outburst interlapsed by
optically thin flares (also see \citealt{Massi2020}).
Optically thin outbursts in microquasars are explained due to shocks and/or magnetic reconnection events
(\citealt{Romero2017} and references therein).

Multiple reconnection events lead to multiple ejection of plasmoids of different size, whose period can vary from minutes, hours to days \citep{Petropoulou2016,Sironi2016}.
On the basis of these models, the magnetic reconnection and consequent formation of plasmoids can occur at two locations in X-ray binaries. 
The first location is magnetic reconnection in the accretion disc \citep{Yuan2009} and the consequent travelling of plasmoids into the jet. 
\citet{Mirabel1998} showed that in \grs{} the ejection of electrons from the accretion disc  were first observed in the X-rays, then in the infrared and finally in the radio band.
These kind of observations, interpreted as accretion disc ejections, are expected during the onset of an outburst and not during its decay. 
Infact, in \citet{Nosel2018}, we studied the location of QPOs along the orbit of \lsi. Statistically, it was found that the X-ray QPOs are present during the onset of the steady radio jet. Possibly those X-ray variability were associated with ejection of plasmoids from the accretion disc.
Other than the accretion disc, the magnetic reconnection events can also occur in the jet itself 
(\citealt{Petropoulou2016, Sironi2016}).
In the jets, radio emission is due to the synchrotron process, with relativistic plasma injected in quasi-periodic fashion.
However, X-ray emission can be either due to synchrotron emission \citep{Markoff2005} or due to up-scattering of the low-energy radio photons via inverse Compton scattering, i.e. synchrotron self-Compton (SSC) (e.g. fig.~3 in \citealt{Romero2017}). 
If the X-ray emission leads the radio emission, it would imply a synchrotron origin and if the radio emission leads the X-ray emission, it would be due to SSC scattering. A similar phenomenon, i.e. the radio emission leading the X-ray emission by about 2~weeks, was observed in the blazar PKS 1510-089 by \citet{marscheretal03} and explained as SSC emission (\citealt{Marscher2010} and references therein). 
In our observations, the found time-lag between the radio and X-ray emission hints that the radio emission leads the X-ray emission but within Gaussian dispersion it could have zero time-lag.
Future sensitive observations covering longer time intervals are clearly mandatory to confirm which physical process dominates at these frequencies.

\section*{Acknowledgements}
The authors would like to thank the referee for their valuable comments and Gisela Ortiz-Leon for carefully reading the manuscript. RS thanks V. M. Pati$\tilde{\textrm{n}}$o-$\Acute{\textrm{A}}$lvarez and C$\Acute{\textrm{e}}$line Chidiac for discussion on DCF.
We thank the staff of the Mullard Radio Astronomy Observatory, University of Cambridge, for their support in the maintenance and operation of AMI. We acknowledge support from the European Research Council under grant ERC-2012-StG-307215 LODESTONE.  Based on observations obtained with \xmm, an ESA science mission with instruments and contributions directly funded by ESA Member States and NASA. The authors acknowledge support by the state of Baden--W\"urttemberg through bwHPC. This work was supported by DFG through the grant MA 7807/2-1.

\section*{Data Availability}
The data underlying this article will be shared on reasonable request to the corresponding author.

%%%%%%%%%%%%%%%%%%%%%%%%%%%%%%%%%%%%%%%%%%%%%%%%%%



\bibliographystyle{mnras}
\bibliography{reference}

%%%%%%%%%%%%%%%%%%%%%%%%%%%%%%%%%%%%%%%%%%%%%%%%%%

%%%%%%%%%%%%%%%%% APPENDICES %%%%%%%%%%%%%%%%%%%%%

%\appendix

%%%%%%%%%%%%%%%%%%%%%%%%%%%%%%%%%%%%%%%%%%%%%%%%%%

% Don't change these lines
\bsp	% typesetting comment
\label{lastpage}
\end{document}